# A Dual-gate MoS₂ Photodetector Based on Interface Coupling Effect

*Fuyou Liao, Jianan Deng, Xinyu Chen, Yin Wang, Xinzhi Zhang, Jian Liu, Hao Zhu, Lin Chen, Qingqing Sun, Weida Hu, Jianlu Wang, Jing Zhou, Peng Zhou, David Wei Zhang, Jing Wan\*, Wenzhong Bao\**

Dr. Fuyou Liao, Dr. Xinyu Chen, Dr. Yin Wang, Xinzhi Zhang, Dr. Hao Zhu, Dr. Lin Chen, Prof. Qinging Sun, Prof. Peng Zhou, Prof. David Wei Zhang, Prof. Wenzhong Bao\*

State Key Laboratory of ASIC and System, School of Microelectronics, Fudan University, Shanghai 200433, China

E-mail: baowz@fudan.edu.cn

Dr. Jianan Deng, Dr. Jian Liu, Prof. Jing Wan\*
E-mail: jingwan@fudan.edu.cn

State Key Laboratory of ASIC and System, School of Information Science and Technology, Fudan University, Shanghai 200433, China

Dr. Weida Hu, Dr. Jianlu Wang, Dr. Jing Zhou,

Shanghai Institute of Technical Physics, Chinese Academy of Sciences, Shanghai 200083, China

**Abstract:**

Two-dimensional (2D) transition metal dichalcogenides (TMDs) based photodetectors have shown great potential for the next generation optoelectronics. However, most of the reported $MoS_2$ photodetectors function under the photogating effect originated from the charge-trap mechanism, which is difficult for quantitative control. Such devices generally suffer from a poor compromise between response speed and responsivity ($R$) and large dark current. Here, a dual-gated (DG) $MoS_2$ phototransistor operating based on the interface coupling effect (ICE) is demonstrated. By simultaneously applying a negative top-gate voltage ($V_{TG}$) and positive back-gate voltage ($V_{BG}$) to the $MoS_2$ channel, the photo-generated holes can be effectively trapped in the depleted region under TG. An ultrahigh $R$ of ~$10^5$ A/W and detectivity ($D^*$) of ~$10^{14}$ Jones have been achieved in several devices with different thickness under $P_{in}$ of 53 μW/cm$^2$ at $V_{TG}$=-5 V. Moreover, the response time of the DG phototransistor can also be modulated based on the ICE. Based on these systematic measurements of $MoS_2$ DG phototransistors, the results show that the ICE plays an important role in the modulation of photoelectric performances. Our results also pave the way for the future optoelectrical application of 2D TMDs materials and prompt for further investigation in the DG structured phototransistors.

2D-TMDs have been rapidly developed in recent years and attracted tremendous research attention in the field of photodetection due to their unique advantages, including broadband response spectrum, tunable bandgap, mechanical flexibility and availability of wafer-scale growth and processing.[1-4] Tremendous efforts have been dedicated to developing high-performance 2D TMDs-based photodetectors for potential applications in optical imaging, sensing and communications.[5-7] Among the big family of TMDs, $MoS_2$ has been widely studied owing to its unique properties. For example, it has a direct bandgap of 1.8 eV as a single-layer and an indirect bandgap of 1.2 eV in a bulk form. [8, 9] Furthermore, it has been reported that the single-layer $MoS_2$ field-effect transistor (FET) has a boost of carrier mobility when integrated with a high-k dielectric layer of $HfO_2$.[10] A monolayer of $MoS_2$ phototransistor has shown a short response time (≈50 ms) and a higher photoresponsivity ($R$) (≈7.5 mA/W) than those of graphene phototransistors.[11] The wavelength dependent $R$ also varies with the thickness of $MoS_2$, exhibiting a wide spectral response range from ultraviolet to infrared. Moreover, the transparent flexible properties are also intriguing towards novel applications such as flexible and wearable electronic devices.[12, 13]

Conventional $MoS_2$ phototransistor relies on the surface trap as photogate modulating the channel current and threshold voltage ($V_{TH}$).[14, 15] Thus, the performances of the device are largely determined by the trap density and energy level. However, these traps are mainly formed by the interface imperfect bond and defects, and thus it is difficult to precisely control their density. This is also a well-known issue in Si-based transistors.[16] Besides, the interface traps might cause persistent photoconductance (PPC) phenomenon, which result in a ultralong decay time after the illumination is switched off.[17-20] Although $MoS_2$ photodetectors already show high responsivity, there is still plenty of room to improve it.[17, 20] It has been reported that $R$ of some MoS$_2$ phototransistors can reach $10^3$-$10^4$ A/W, which was achieved under extreme conditions, and at the expense of large dark current,[20, 21] which limit the practical photodetection applications. Some other approaches which have been reported to enhance the photoresponse of MoS$_2$ photodetectors. For example, Kufer $et$ $al.$[22] fabricated a HfO$_2$ encapsulated MoS$_2$ photodetector that exhibits an $R$ of ~$10^4$ A/W at $V_D$=5 V. Huo $et$ $al.$[23] fabricated out-of-plane MoS$_2$ p-n homojunction through chemical doping method and report $R$ of 7×$10^4$ A/W under an extremely low illumination power density ($P_{in}$) of 73 pW/cm$^2$ at $V_D$=10 V and $V_G$=60 V. Wu $et$ $al.$[24] developed a MoS$_2$ photodetector fabricated on the Al$_2$O$_3$/ITO /SiO$_2$/Si substrate with an $R$ of 2.7×$10^4$ A/W. However, these approaches still work by the mechanism of photo-carriers trapped in surface states, and suffer from the shortcoming of poor compromise between response speed and responsivity. A summary of figures-of-merit for previous reported MoS$_2$ based photodetectors is seen table S1. It is of great importance to further explore other new detecting mechanisms to realize both high response speed and responsivity based on MoS$_2$ phototransistors.

In this work, in order to improve and control the $R$ and response speed of the MoS$_2$ based photodetector simultaneously, a dual-gated (DG) architecture is utilized by integrating both top and back metal gates with the high-$k$ dielectric layer. Here, in our proposed DG MoS$_2$ phototransistor, the back gate (BG) is positively biased in order to form an electron channel at the back interface. Meanwhile, the local top gate (TG) is negatively biased. When the device is under illumination, the photo-generated holes accumulate just beneath the TG. These accumulated photo holes can effectively screen the electric field from the TG, and lead to a boost of the drive current compared with that in dark condition. Such an interface coupling effect (ICE) enables our device with outstanding photoelectric characteristics. The experimental results show that this DG MoS$_2$ phototransistor with a thickness between 2.5 and 6.5 nm yields ultrahigh $R$ of $\sim 10^5$ A/W and ultrahigh detectivity ($D^*$) of $\sim 10^{14}$ Jones under $P_{in}$ of 53 μW/cm$^2$. Furthermore, the responsivity $R$ and response time $t_{rise}$ or $t_{fall}$ can be efficiently tuned by $V_{TG}$. Our results indicated that the DG MoS$_2$ phototransistor would be a great potential candidate for the next-generation optoelectronic devices.

**Figure 1**a displays a three-dimensional (3D) schematic of DG MoS$_2$ phototransistor on Si/SiO$_2$ substrate with two metal gates located on the top and bottom of the MoS$_2$ layer, respectively. The non-gated regions (underlap) for TG are engineered to keep a sufficient total illumination reception area, which is critical to improving the light absorption efficiency. A detailed device fabrication procedure is described in the Methods section. The optical microscope of the device is shown in Figure 1b. The height profile and atomic force microscope (AFM) image of the multilayer MoS$_2$ are shown in Figure 1c. The thickness of MoS$_2$ in this device is measured by about 5.6 nm, suggesting an 8-layer MoS$_2$ channel. It should be mentioned here that we have fabricated a batch of DG MoS$_2$ phototransistors with varied thickness. Next, we measured the basic electrical characteristics of DG MoS$_2$ phototransistor under dark conditions. The output and transfer curves of BG mode ($V_{TG}$=0 V) and TG mode ($V_{BG}$=0 V) of this device are measured, see Figure S1. The transfer characteristics are measured by sweeping the $V_{BG}$ with different $V_{TG}$ from -5 to +3 V at a step of 1 V and $V_D$=0.1 V, as shown in Figure 1d. It is worth noting that negative $V_{TG}$ shifts the transfer curves toward a positive direction, whereas for a positive $V_{TG}$, the transfer curve hardly shifts in the opposite direction,

which indicates that negative $V_{TG}$ can effectively influence the channel current due to the global BG and short local TG architecture in this device. The $V_{TH}$ value of the DG $MoS_2$ FET with different $V_{TG}$ is calculated by extrapolating the linear portion of $I_D$-$V_{BG}$ curve to a zero drain current point. It is found that the DG $MoS_2$ FET has a large $V_{TH}$ modulation from -0.73 to 0.22 V when the $V_{TG}$ is changed from +3 to -5 V, which indicates a strong interface coupling between the BG and TG.

To investigate the photoelectric characteristics of the DG $MoS_2$ phototransistor, we measured the transfer curves of the devices by sweeping $V_{BG}$ with different $V_{TG}$ under dark and a $P_{in}$ of 1.55 $mW/cm^2$ (550 nm). As shown in **Figure 2**a, a negative $V_{TG}$ can greatly suppress the dark current in the $MoS_2$ phototransistor, and when the device biased at $V_D$=1 V is illuminated by a focused laser beam, the transfer curves shift negatively, which indicates an obvious boost of the photocurrent due to the photogating effect. The corresponding output curves with varied $V_{TG}$ are also measured (see Figure S2). For comparison the performance of DG and BG $MoS_2$ phototransistor based on the same process and bias condition, the same measurement is performed on BG-FET (raw data see Figure S3). In order to quantitatively compare the photogating effect in the DG $MoS_2$ phototransistor under different $V_{TG}$, the change of $V_{TH}$ ($\Delta V_{TH}$) is calculated by $\Delta V_{TH} = \left| V_{TH-illumination} - V_{TH-dark} \right|$. The result as shown in Figure 2b indicates that the $\Delta V_{TH}$ increase from 0.1 to 0.5 V as $V_{TG}$ decreasing from +3 to -6 V, and the $\Delta V_{TH}$ of BG-FET is ~0.1 V, which means the photogating effect is stronger when $V_{TG}$ is more negative. One of the most important figures of merit for phototransistor is $R$. According to Figure 2a, $R$ of the DG $MoS_2$ phototransistor is calculated by $R = \dfrac{I_{ph}}{P_{in} \times A}$, where $I_{ph}$ is the photocurrent, $P_{in}$ is the illumination power density and $A$ is the active area. Figure 2c shows the variation of $R$ as a function of $V_{BG}$ under different $V_{TG}$ biases at $V_D$=1 V. For a fixed $V_{TG}$, $R$ rises while $V_{BG}$ increases from a negative value to positive value, because increasing $V_{BG}$ not only enhances the ICE but also increases the transconductance ($g_m$). $R$ reaches the peak value and then drops when the $V_{BG}$ continues increasing due to the decrease of $g_m$, as $R$ depends on $g_m$ when the $\Delta V_{TH}$ is constant. The maximum $R$ value ($R_{max}$) increases as the $V_{TG}$ decreases from +3 to -6 V. $R_{max}$ under varied $V_{TG}$, as well as the corresponding $D^*$ are extracted and as shown in Figure 2d. Assuming that noise from dark current is the major factor, $D^*$ can be calculated by $D^* = \sqrt{A / 2eI_D} \times R$, where $A$ is the active area of the detector in a unit of cm$^2$, $e$ is the electronic charge, $I_D$ is the dark current and $R$ is the responsivity. The results clearly show that $R_{max}$ is $4.2 \times 10^4$ A/W when $V_{TG}$=-6 V and drops to $1.1 \times 10^4$ A/W when $V_{TG}$ increases to -1 V and then remains nearly unchanged when $V_{TG}$ increases to +3 V, whereas the $R_{max}$ of BG-FET is only $7 \times 10^3$ A/W. The $D^*$ dependence of $V_{TG}$ curve exhibits the same trend as the $R_{max}$ -$V_{TG}$ relation curve. The maximum $D^*$ is $9.6 \times 10^{12}$ Jones at $V_{TG}$=-6 V and the minimum $D^*$ is $3.2 \times 10^{12}$ Jones when $V_{TG}$ =-1 V, which is much higher than the $D^*$ of BG-FET ($9.42 \times 10^{11}$ Jones). The results demonstrate that the DG MoS$_2$ phototransistor can achieve higher $R$ and $D^*$ when $V_{TG}$ becomes lower due to the ICE and much enhancement compare to those of BG-FET. A detailed photoresponse characteristic of DG phototransistor with different MoS$_2$ thickness ranging from 1.6 to 8 nm are shown in Figure S4-7, these devices with a thickness between 2.5 and 6.5 nm show a strong ICE when applied negative $V_{TG}$ and positive $V_{BG}$. The ultrahigh $R$ and $D^*$ values are superior to those of most reported devices, such as MoS$_2$ photodetector encapsulated by HfO$_2$ ($R$=10$^4$ A/W, $D^*$=7.7$\times$10$^{11}$ Jones)[22] and substrate-enhanced MoS$_2$ photodetector ($R$=10$^4$ A/W, $D^*$=6.4$\times$10$^{11}$ Jones).[25] It should be noted that $D^*$ of photodetectors based on 2D materials can be extracted by two different methods, dark current or low frequency noise measurement. To get a fast estimation of the detectivity, $D^*$ is often extracted by assuming that the main noise is from the shot noise of the dark current as $S_n$=(2$eI_{dark}$)$^{1/2}$.[26-30] However, this method neglects other noise components, such as 1/$f$ noise which is typically seen in a MOSFET-based phototransistor. In order to obtain a complete assessment of $D^*$, low frequency noise measurement is also used to extract the $D*$ of our device as $D* = R \times \dfrac{\sqrt{AB}}{i_N}$, where $B$

is the bandwidth, and $i_N$ the measured noise current of the device. [22]

As shown in Figure S8a, the device indeed shows $1/f$ noise, similar to other MOSFET devices.[31-33] Figure S8b shows that the noise reduces as the $V_{BG}$ decreases due to the lower drain current. This leads to an extracted $D*$ up to $5 \times 10^{11}$ Jones at $V_{BG} = 1.5$ V, which is about 2 orders lower than that extracted directly from the dark current, see Figure S8c. This is mainly due to the $1/f$ noise component which is not considered in the extraction by dark current. In order to fairly compare our results with other publications, table S1 has compared our results to other publications with $D*$ extracted both from dark current and low frequency noise. We have also labeled the published results with yellow background to denote $D*$ extracted by low frequency noise measurements. As can be seen from the comparison table, our device shows very competent $D*$ values from both dark current and low frequency noise measurement

The observed ICE of the DG MoS$_2$ phototransistor can be explained by a simple schematic illustration (Figure 2e-f). When the DG MoS$_2$ phototransistor works in the condition of $V_{BG}$>0 V and $V_{TG} \geq 0$ V, $V_{TG}$ just increases the electron concentration beneath the TG. Under illumination, the absorption of photons generates electron-hole pairs and the holes are then trapped in the localized states near the valence band edge, which is similar to that in the conventional BG MoS$_2$ phototransistor. In this case, the trapped holes induce more electrons through electric field induction. Thus enhanced electron density lowers the total resistance of the device, resulting in a larger current flow ($I_{ph}$) so that the transfer curve is horizontally shifted with respect to the dark transfer curve due to the trapped photoholes in the surface, as shown in Figure 2e. However, when a negative $V_{TG}$ ($V_{TG}$<0 V) is applied on the TG with $V_{BG}$>0 V, the electron channel under TG is depleted through the ICE, and thus reduces the $I_D$ flowing at the back interface. As shown in Figure 2f, the light illumination generates extra holes that gather close to the top interface due to $V_{TG}$<0 V and screen the electrical field induced by $V_{TG}$, thereby diminishing the interface coupling, and consequently the electron current at the bottom interface is recovered. The operation of the DG MoS$_2$ phototransistor is also similar to that of an n-type JFET. The hole layer or electron-depleted layer induced by $V_{TG}$ at the top interface functions as the JFET gate controlling the bottom channel induced by $V_{BG}$. The photogenerated holes accumulating at the top interface increase the potential of the field-induces JFET's gate, which is similar with applying a positive $V_{BG}$ to JFET that would increase the electron current in the bottom channel. It should be noted that the MoS$_2$ channel thickness of DG-FET needs to be in an appropriate range to obtain a strong photoresponse enhancement when applying a negative $V_{TG}$. Because negative $V_{TG}$ can turn off a thin MoS$_2$ FET which results in suppression of photocurrent. In contrast, there is negligible ICE when the MoS$_2$ channel is too thick, because a negative $V_{TG}$ has little impact on the current at the bottom interface. This mechanism leads to the phenomenon of obvious ICE in the DG phototransistors with MoS$_2$ thickness from 2.5 to 6.5 nm, which will be further discussed below. Although the ICE has been reported in SOI photodetector, [34] it is first reported in the DG MoS$_2$ phototransistor. It is worth noting that there are both ICE and states trap in the DG device with negative $V_{TG}$, but the former plays a more important role in the DG MoS$_2$ phototransistor due to the strong top gate controllability.

We further investigate the dominant photocurrent generation mechanism of the DG MoS$_2$ phototransistor under different incident light power densities. **Figure 3**a-b exhibits the linearly scaled output curves and transfer curves with $V_{TG}$=-5 V and $P_{in}$ increasing from 0 to 1.55 mW/cm$^2$. The transfer curves with $V_{TG}$=0 and +3 V under the same condition are shown in Figure S9. The corresponding output curves with varying $V_{BG}$ at $V_{TG}$=-5, 0 and 3 V are shown in Figure S10. It is clearly observed that the $I_D$ increases and $V_{TH}$ shifts negatively with an increase of $P_{in}$, which behaves like a typical photogating effect. Under illumination with negative $V_{TG}$, the photo-generated holes accumulate in the region under TG are captured by the TG electric filed. The accumulated holes thus screen the field from TG, leading to a recovery of the $I_D$. At the same time, the captured holes also act as an additional positive TG voltage to tune the $I_D$. Both factors cause an increase of electron concentration in the $MoS_2$ channel, which makes the channel more n-type. This corresponds to the transfer characteristics in which the $V_{TH}$ moves negatively with increasing $P_{in}$. In order to compare the photogating effect originated from such interface coupling and trap states in our device, we extracted the $\Delta V_{TH}$ as a function of $P_{in}$ under $V_{TG}$ = -5, 0 and +3 V, respectively, as shown in Figure 3c, and $\Delta V_{TH}$ of BG-FET is also extracted for comparison. It clearly shows that $\Delta V_{TH}$ of DG-FET at $V_{TG}$=-5 V is much higher than that of $V_{TG}$=0 and =3 V, and that of BG-FET, indicating that the interface coupling mechanism is dominant against the trap-related photogating effect when applying $V_{TG}$=-5 V to the DG $MoS_2$ device.

The dependence of $I_{ph}$ on $P_{in}$ at different $V_{BG}$ is shown in Figure 3d. We found that $I_{ph}$ increases linearly with $P_{in}$ ( $I_{ph} \propto P_{in}^{\alpha}$, $\alpha \approx 1$) at negative BG bias ($V_{BG}$=-1.5 V), which indicates that the photo-generated carriers are positively proportional to the total phonon flux, similar to the mechanism of a photoconductive detector. However, as $V_{BG}$ increases and the $MoS_2$ channel is turned on, the correlation between $I_{ph}$ and $P_{in}$ becomes sublinear ($\alpha \approx 0.3$ at $V_{BG}$=1.5 V), it indicates that the interface coupling associated with the trap states plays a more important role. On the other hand, under a low illumination power, most of the incident light can be absorbed by the $MoS_2$, while more is wasted when the $P_{in}$ is high because of the light absorption saturation. [35] Figure 3e shows the correlation of $R$ and $P_{in}$ under different $V_{BG}$ values, where $R$ reaches an ultrahigh value of $3.6 \times 10^5$ A/W at $V_D$=1 V and $V_{BG}$=1.5 V under illumination power intensity of 53 $\mu$W/cm$^2$. The corresponding $P_{in}$ dependence of $R$ with different thickness of MoS$_2$ from 1.6 to 8 nm are also measured (see Figure S11 for details). In Figure 3f, the thickness dependence of $R$ and $D^*$ ($V_{TG}$=-5 V, $V_D$=1 V) under illumination of 53 $\mu$W/cm$^2$ and 1.55 mW/cm$^2$ are both extracted, respectively. A large $R$ in the order of $10^4$ and $10^5$ A/W can be obtained and is nearly independent on the thickness between 2.5 and 6.5 nm. However, there is a downward trend as the thickness of MoS$_2$ is thinner than 2.5 nm or thicker than 6.5 nm. With too thin film, the top photohole and bottom electron layers are very close, leading to high recombination in the channel. This high carrier recombination degrades the responsivity. With too thick MoS$_2$ film, the ICE becomes weak and thus leading to degraded responsivity as well. In addition, $D^*$ is extracted to be $\sim 10^{13}$ and $\sim 10^{14}$ Jones at the range of 1.6 to 8 nm, under $P_{in}$ of 1.55 mW/cm$^2$ and 53 $\mu$W/cm$^2$, respectively, which demonstrates an excellent enhancement of MoS$_2$ phototransistor performance.

**Figure 4**a compares the normalized transient characteristics of the DG phototransistor with the conventional single (back gated) BG phototransistor at $V_{BG}$=-3 V based on the 5.6 nm MoS$_2$ flake. The photocurrent of both devices increases when the illumination is turned on and decreases after the illumination is turned off. The BG MoS$_2$ phototransistor, like conventional MoS$_2$ phototransistor reported in the references, shows a slow $I_{ph}$ rise and a very slow $I_{ph}$ decrease due to the PPC. This PPC phenomenon of MoS$_2$ phototransistor was also previously reported by several groups [18-20] and observed in other semiconductors. [36-38] In contrast, the DG MoS$_2$ phototransistor with negative $V_{TG}$ has much faster photoresponse than the BG phototransistor due to the interface coupling mechanism, indicating the obvious advantage of this device in response speed (raw data of photocurrent pulses of the DG device with various

$V_{BG}$ at fixed $V_{TG}$ of -5 V at 1.55 mW/cm$^2$ see Figure S12). Figure 4b shows time-resolved $I_{ph}$ under different values of $P_{in}$ for the wavelength of 550 nm (left), and different wavelength with $P_{in}$ of ~1.55 mW/cm$^2$ (right) at $V_{BG}$=+2 V and $V_{TG}$=-5 V. It is noteworthy that corresponding $I_{ph}$ increases with the increase of $P_{in}$ or with the decrease of wavelength. For example, the $I_{ph}$ increasing from 4.1 to 14.7 μA as the $P_{in}$ changes from 82 μW/cm$^2$ to 1.55 mW/cm$^2$ for the wavelength of 532 nm at $V_D$ = 1 V. When the wavelength decreases from 650 to 350 nm, it leads to an obvious increase of $I_{ph}$ from 5.6 to 35.4 μA. A higher $P_{in}$ or a shorter wavelength corresponds to more photons or higher photon energies, hence more electron-hole pairs can be generated and a larger $I_{ph}$ can be achieved. Figure 4c shows the time-resolved $I_{ph}$ for the DG MoS$_2$ phototransistor with $V_{BG}$=+2 V, $V_{TG}$=-4, -5 and -6 V, respectively. It is clear that when $V_{TG}$ decreases from -4 to -6 V, the $I_{ph}$ increases from 13 to 16.5 μA, which is consistent with the above conclusion. Furthermore, the response speed of DG MoS$_2$ phototransistor can be tuned by different $V_{TG}$, as shown in Figure 4d. The rise time ($t_{rise}$) is defined as the time for $I_{ph}$ to increase from 10 to 90%, and vice versa for the fall time ($t_{fall}$).[14] When $V_{TG}$ decreases from -4 to -6 V, the $t_{rise}$ decreases from 11.9 s to 8.3 s and $t_{fall}$ decreases from 76.2 s to 46.4 s. Thus, it confirms that $V_{TG}$ plays an important role in modulating the response time, especially the $t_{fall}$. Although the response speed of MoS$_2$ DG-FET is much faster than that of the BG-FET, it is still not qualified for practical application. It can be explained that the charging and discharging time of the top gate capacitor limits the further increase in speed. There are several methods to further improve the response speed, such as applying a more negative $V_{TG}$ based on a stronger ICE, reducing the top gate capacitance by shortening the front gate length and applying a smaller $V_{BG}$ at the expense of $R$.

In summary, we have demonstrated a DG MoS$_2$ phototransistor with ultrahigh $R$ of ~10$^5$

A/W and ultrahigh $D^*$ of ~$10^{14}$ Jones under $P_{in}$ of 53 μW/cm$^2$ and a fixed gate voltage combination ($V_{BG}$=2 V, $V_{TG}$=-5 V). It is found that the DG MoS$_2$ phototransistor can achieve a better compromise between response speed and responsivity than the conventional BG phototransistor due to the ICE. By controlling the $V_{TG}$, the responsivity and temporal response can be effectively manipulated. The thickness-dependent photoresponsivity in DG MoS$_2$ phototransistors has also been investigated and all devices exhibit ultrahigh $R$ and $D^*$. The DG device architecture based on interface coupling provides a new route for improving the performance of TMD based photodetectors.

**Experimental Section**

*Device Fabrication:* The device fabrication begins with patterning a local BG on a heavily doped p-type Si substrate capping with a layer of 300-nm-thick thermal oxide. Dry and wet etching is then carried out to etch a 40-nm-thick trench on SiO$_2$, followed by deposition and lift-off of Ti/Au (10 nm/30 nm) to fill the trench using E-beam evaporation. Next, a layer of 15 nm HfO$_2$ is deposited by atomic layer deposition (ALD) at 180 °C to form the BG dielectric. The multilayer MoS$_2$ flakes are obtained by using the micromechanical exfoliation method,[39] and then precisely transferred onto the top of the buried BG electrodes using a polydimethylsiloxane (PDMS) stamp and customer designed aligner. Ti/Au (10 nm/ 40 nm) electrodes are deposited to form source and drain contacts, and the channel length between the contacts is 4 μm. Then 2 h high vacuum annealing is carried out at 200 °C to improve the metal-MoS$_2$ interface. Next, a 2-nm-thick Al$_2$O$_3$ layer is deposited as the seeding layer [40] for subsequent ALD deposition of 15 nm HfO$_2$ and 30 nm Au TG electrode.

*Material and Device Characterizations:* The thickness of MoS$_2$ sheets were measured by atomic force microscopy (AFM, Bruker Dimension Edge). The electrical characteristics of the DG phototransistor were measured using a semiconductor analyzer (Agilent B1500A). Optoelectronic measurement was performed using a xenon lamp as the light source. The white light is filtered with a monochromator and then passes through an optical fiber to illuminate the as-fabricated devices. A light intensity meter was used to calibrate the illumination intensity. All measurements were carried out in air at room temperature.

**Supporting Information**

Supporting Information is available from the Wiley Online Library or from the author.

**Acknowledgments**

F. L. and J. D. contributed equally to this work. This work was supported by the National Key Research and Development Program (2016YFA0203900), National Natural Science Foundation of China (61904032), Shanghai Municipal Science and Technology Commission (18JC1410300) and National Natural Science Foundation of China (61874154).

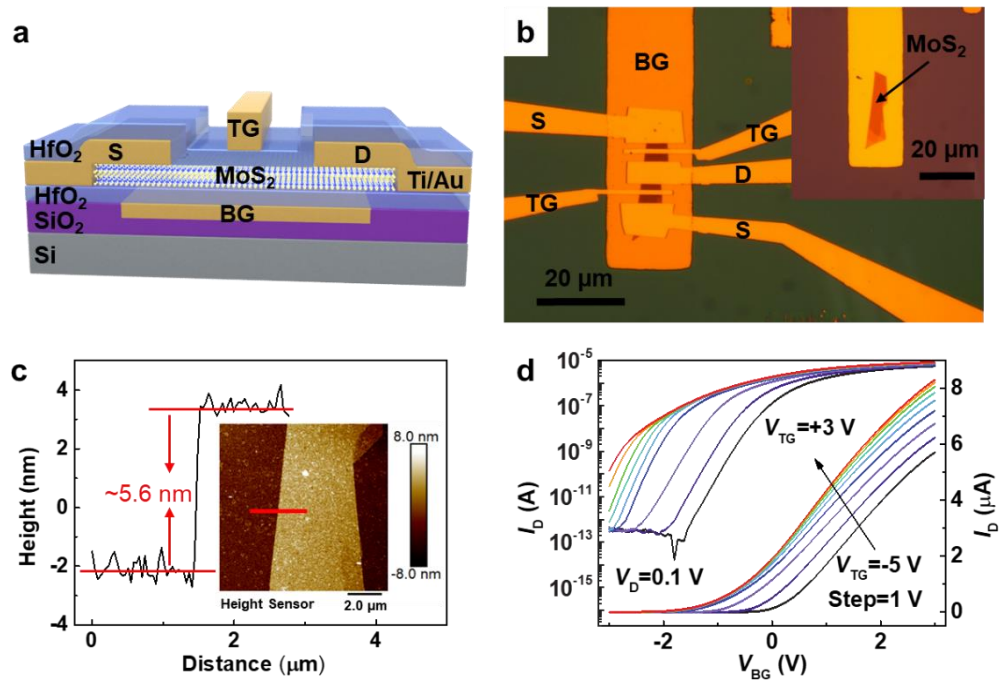

**Figure 1**. A DG MoS$_2$ phototransistor. a) 3D schematic figure of a DG MoS$_2$ phototransistor. b) Optical microscopy image of the DG MoS$_2$ phototransistor. Inset: Optical microscopy image of the multilayer MoS$_2$ flake after transfer to the buried BG. The scale bar is 20 μm. c) Thickness scan along the red line across the boundary of the multilayer MoS$_2$ flake. Inset: AFM image of the multilayer MoS$_2$ flake. d) Transfer characteristics of the DG MoS$_2$ phototransistor by sweeping $V_{BG}$ with $V_D$=0.1 V and varied $V_{TG}$ under dark conditions.

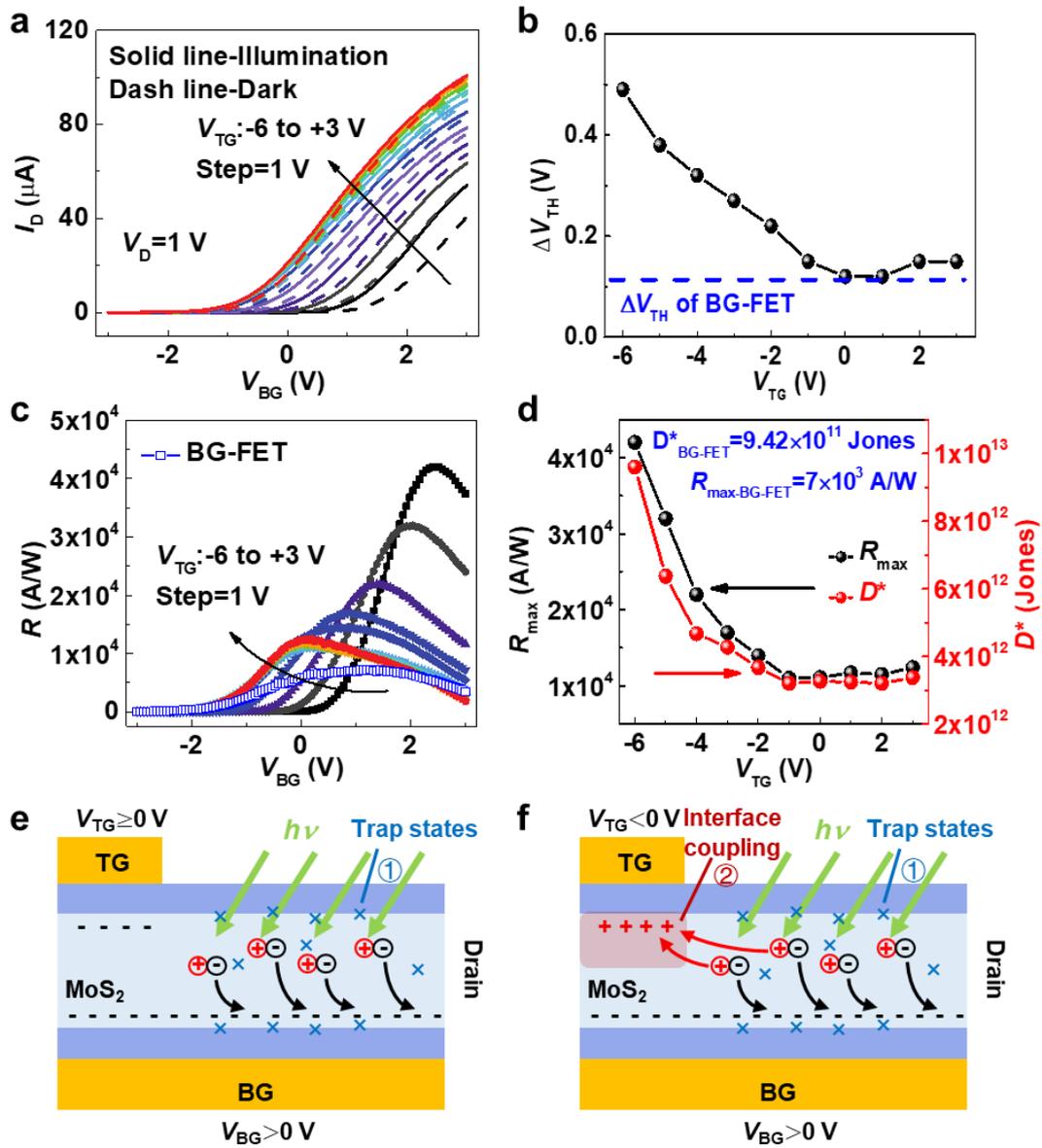

**Figure 2.** Photoresponse characteristic of DG MoS$_2$ phototransistor with different $V_{TG}$. $P_{in}$ is 1.55 mW/cm$^2$ with a wavelength of 550 nm. a) Transfer characteristics of the devices by sweeping $V_{BG}$ under varied $V_{TG}$ from -6 to +3 V at a step of 1 V under dark (dash line) and illumination (solid line) condition. b) $V_{TH}$ shift under different $V_{TG}$ due to illumination. Blue dash line is $\Delta V_{TH}$ of BG-FET. c) $R$ as a function of $V_{BG}$ under different $V_{TG}$. Blueline + square symbol is $R$ of BG-FET. d) The corresponding $R_{max}$ and $D^*$ as a function of $V_{TG}$. e-f) Schematic illustration of generated electron-hole pairs in the DG MoS$_2$ phototransistor with positive (e) and negative (f) $V_{TG}$ biases, respectively.

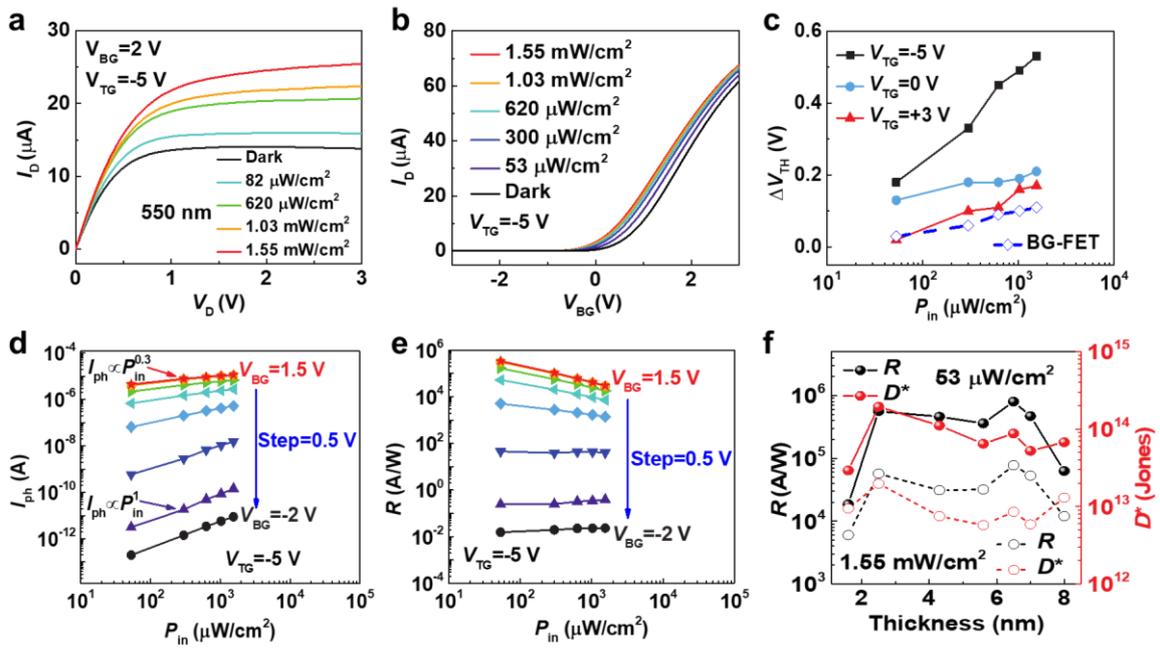

**Figure 3.** $P_{in}$ dependence of the DG MoS$_2$ phototransistor under the illumination of 550 nm with V$_D$=1 V and $V_{TG}$=-5 V. a-b) Output curves and transfer curves for dark condition and illumination with different $P_{in}$. c-e) The $P_{in}$ dependence of $\Delta V_{TH}$, $I_{ph}$, and $R$, respectively. f) The MoS$_2$ thickness dependence of $R$ and $D^*$ under the illumination of 53 μW/cm$^2$ (Solid sphere) and 1.55 mW/cm$^2$ (open circle).

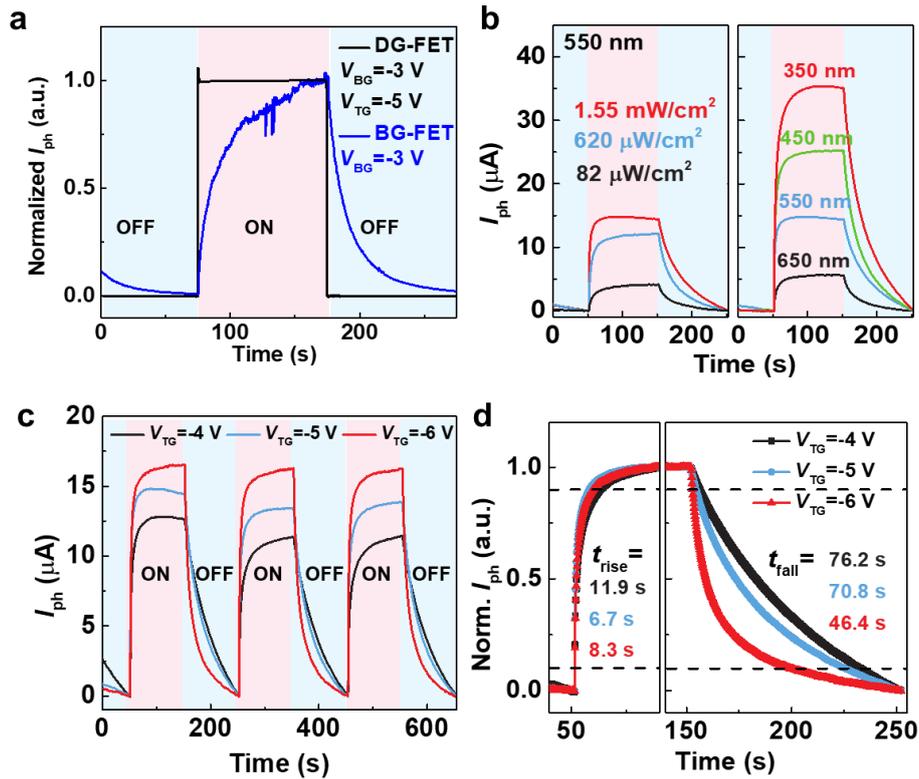

**Figure 4.** Photoswitching characteristics of the DG MoS$_2$ phototransistor. a) Normalized $I_{ph}$ of DG and BG MoS$_2$ phototransistor based on the same thickness flake and operating with the same $V_{BG}$ of -3 V under the illumination of 1.55 mW/cm$^2$ (550 nm). b) Time-resolved $I_{ph}$ for the DG MoS$_2$ phototransistor under different illumination power intensity (left) and different wavelength (right) at $V_{BG}$=+2 V and $V_{TG}$=-5 V. c) Time-resolved $I_{ph}$ for the DG MoS$_2$ phototransistor with $V_{BG}$=+2 V, $V_{TG}$=-4, -5 and -6 V, respectively (550 nm, 1.55 mW/cm$^2$). d) The rise (left) and fall (right) of the normalized $I_{ph}$ with varied $V_{TG}$ extracted from (c). All the $V_D$ is 1 V.

A dual-gated multilayer MoS$_2$ phototransistor is fabricated to demonstrate an interface coupling effect (ICE) for optoelectronic applications. Various device performance parameters can be modulated based on the ICE. An ultrahigh photoresponsivity of ~10$^5$ A/W and detectivity of ~10$^{14}$ Jones have been achieved.

**Photodetectors**

Fuyou Liao, Jianan Deng, Xinyu Chen, Yin Wang, Xinzhi Zhang, Jian Liu, Hao Zhu, Lin Chen, Qingqing Sun, Weida Hu, Jianlu Wang, Jing Zhou, Peng Zhou, David Wei Zhang, Jing Wan*, Wenzhong Bao*

**A Dual-gate MoS₂ Photodetector Based on Interface Coupling Effect**

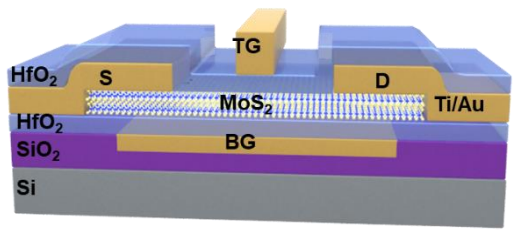

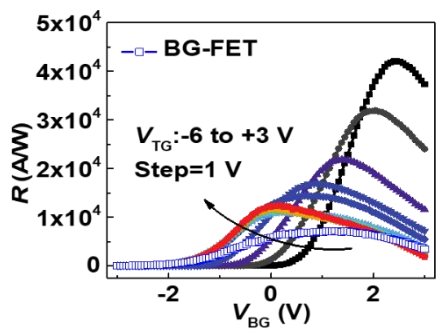

## Supporting Information

**A Dual-gate MoS2 Photodetector Based on Interface Coupling Effect**

*Fuyou Liao, Jianan Deng, Xinyu Chen, Yin Wang, Xinzhi Zhang, Jian Liu, Hao Zhu, Lin Chen, Qingqing Sun, Weida Hu, Jianlu Wang, Jing Zhou, Peng Zhou, David Wei Zhang, Jing Wan\*, Wenzhong Bao\**

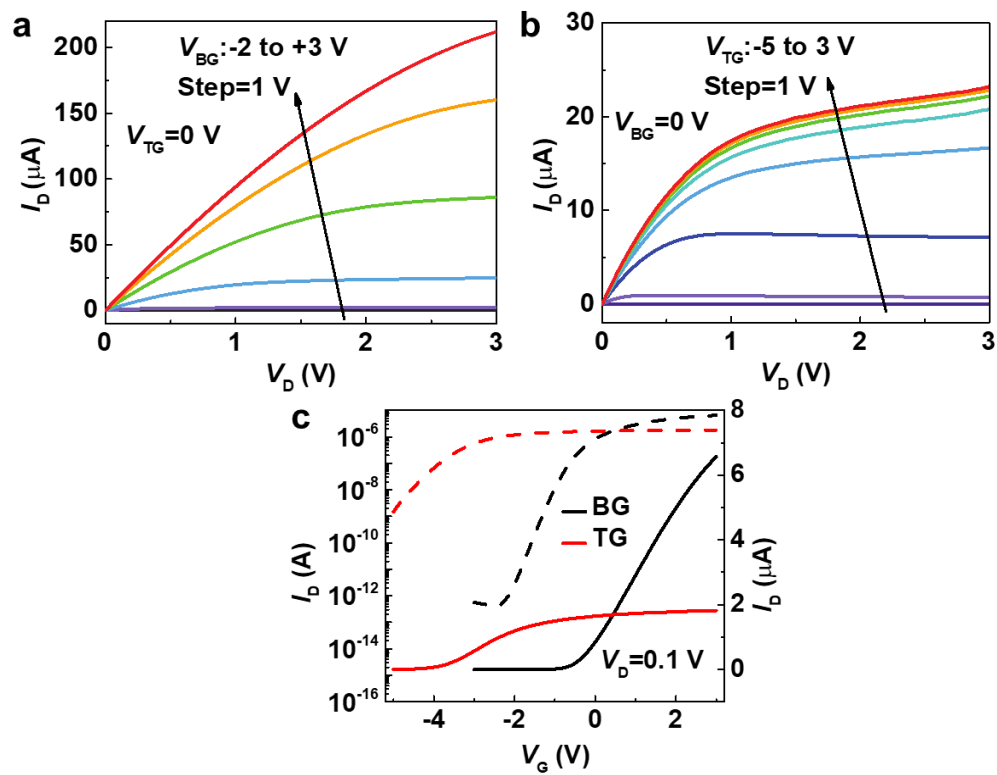

**Figure S1.** (a) $I_D$-$V_D$ curves with $V_{TG}$=0 V and $V_{BG}$ ranging from -2 to +3 V (BG mode). (b) $I_D$-$V_D$ curves with $V_{BG}$=0 V and $V_{TG}$ ranging from -5 to +3 V (TG mode). (c) The BG (black line) and TG (red line) transfer curves of the DG MoS2 phototransistor with $V_D$=0.1 V. The measurement is in ambient air and in the dark.

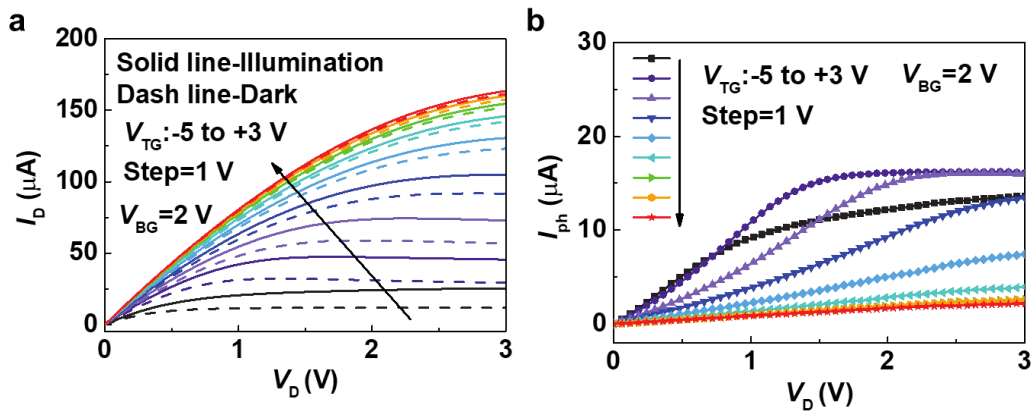

**Figure S2.** (a) The $I_D$-$V_D$ curves of the 5.6 nm MoS$_2$ DG phototransistor with $V_{BG}$=+2 V and $V_{TG}$ ranging from -5 to +3 V in dark and under illumination. $P_{in}$ is 1.55 mW/cm$^2$ with a wavelength of 550 nm. (b) The corresponding $I_{ph}$ is extracted from (a).

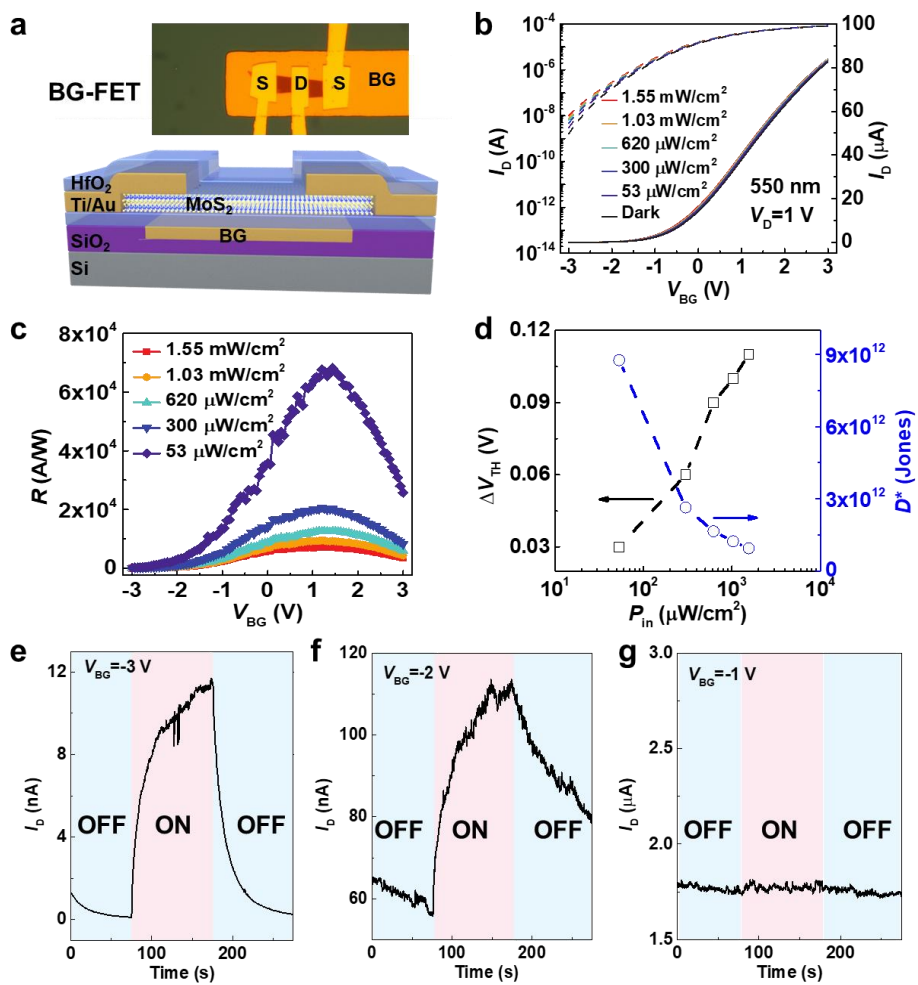

**Figure S3.** A 5.6 nm MoS$_2$ bottom-gate phototransistor (a) and its corresponding optoelectronic response (b-d) and photoswitching characteristics (e-g).

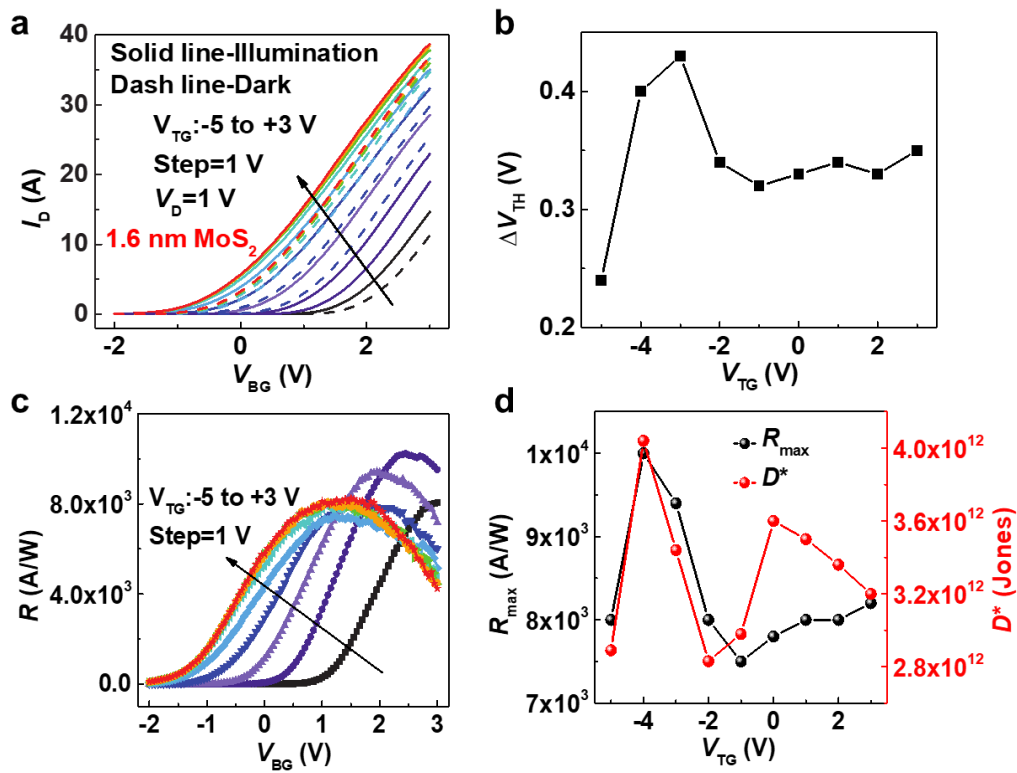

**Figure S4.** Photoresponse characteristic of a 1.6 nm MoS$_2$ DG phototransistor with different $V_{TG}$. $P_{in}$ is 1.55 mW/cm$^2$ with a wavelength of 550 nm. (a) Transfer characteristics of the devices by sweeping $V_{BG}$ under varied $V_{TG}$ from -5 to +3 V at a step of 1 V under dark (dash line) and illumination (solid line) condition. (b) $V_{TH}$ shift under different $V_{TG}$ due to illumination. (c) $R$ as a function of $V_{BG}$ under different $V_{TG}$. (d) The corresponding $R_{max}$ and $D^*$ as a function of $V_{TG}$.

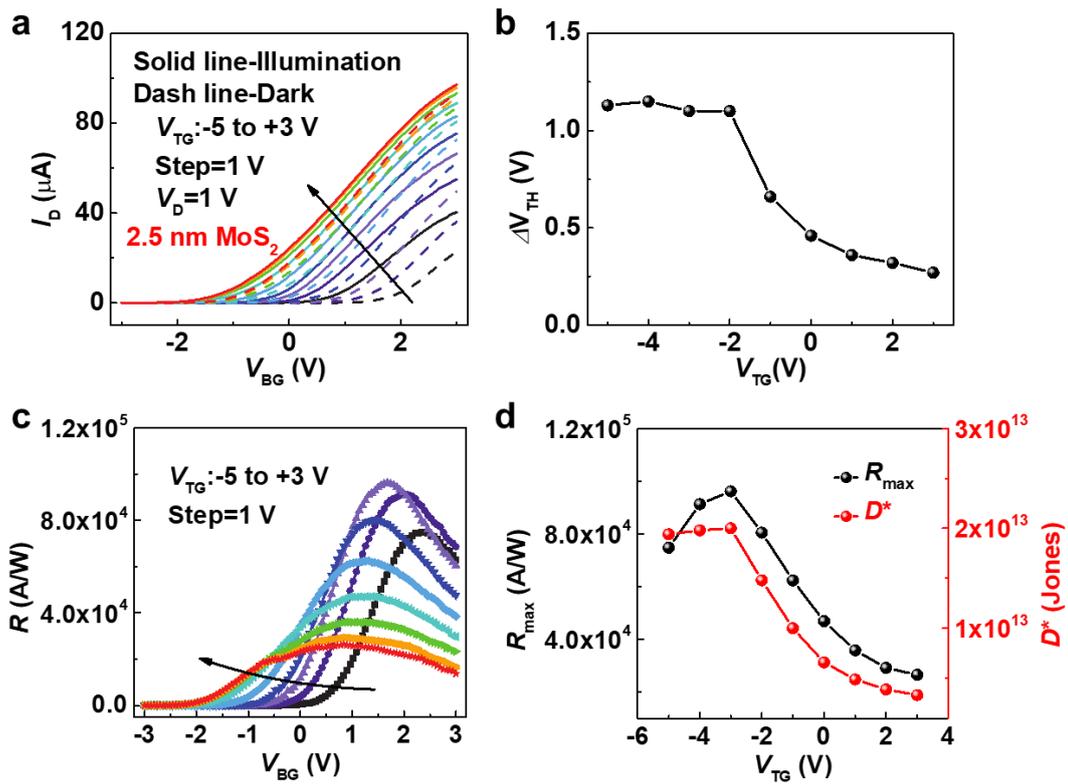

**Figure S5.** Photoresponse characteristic of a 2.5 nm MoS$_2$ DG phototransistor with different $V_{TG}$. $P_{in}$ is 1.55 mW/cm$^2$ with a wavelength of 550 nm. (a) Transfer characteristics of the devices by sweeping $V_{BG}$ under varied $V_{TG}$ from -5 to +3 V at a step of 1 V under dark (dash line) and illumination (solid line) condition. (b) $V_{TH}$ shift under different $V_{TG}$ due to illumination. (c) $R$ as a function of $V_{BG}$ under different $V_{TG}$. (d) The corresponding $R_{max}$ and $D^*$ as a function of $V_{TG}$.

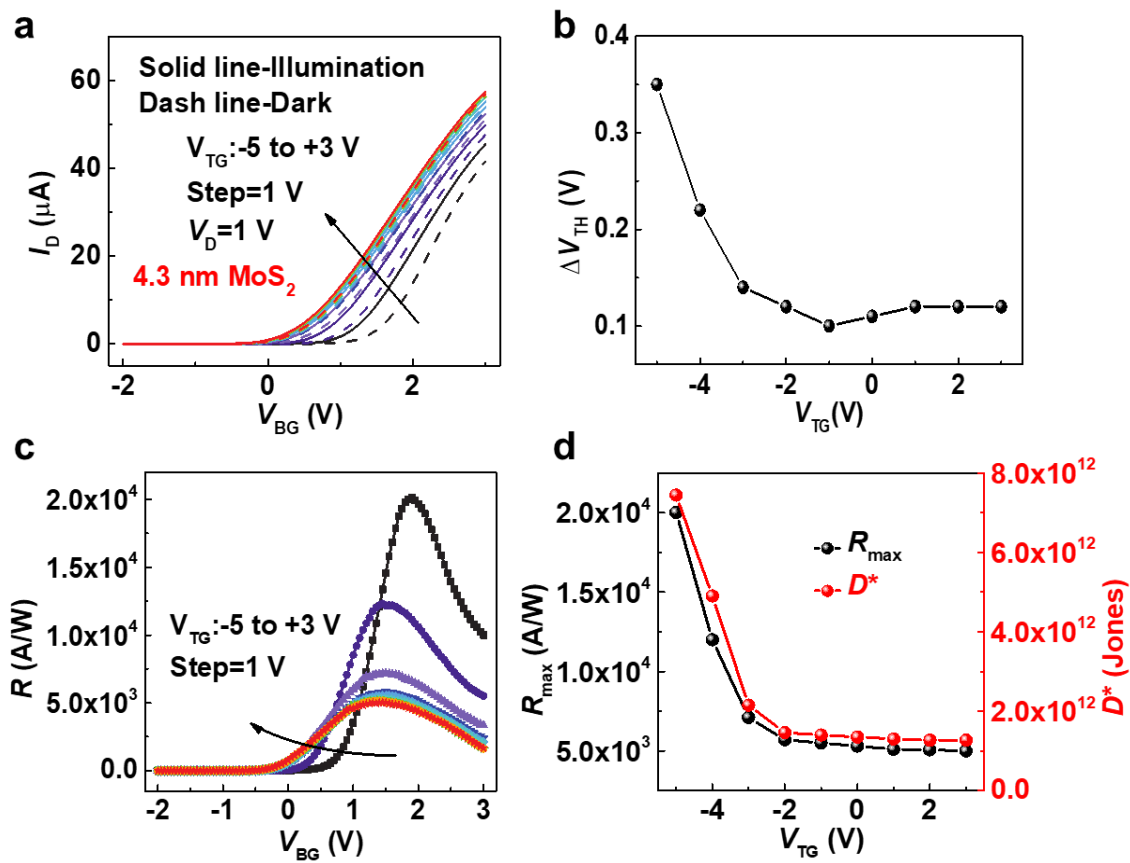

**Figure S6.** Photoresponse characteristic of a 4.3 nm MoS$_2$ DG phototransistor with different $V_{TG}$. $P_{in}$ is 1.55 mW/cm$^2$ with a wavelength of 550 nm. (a) Transfer characteristics of the devices by sweeping $V_{BG}$ under varied $V_{TG}$ from -5 to +3 V at a step of 1 V under dark (dash line) and illumination (solid line) condition. (b) $V_{TH}$ shift under different $V_{TG}$ due to illumination. (c) $R$ as a function of $V_{BG}$ under different $V_{TG}$. (d) The corresponding $R_{max}$ and $D^*$ as a function of $V_{TG}$.

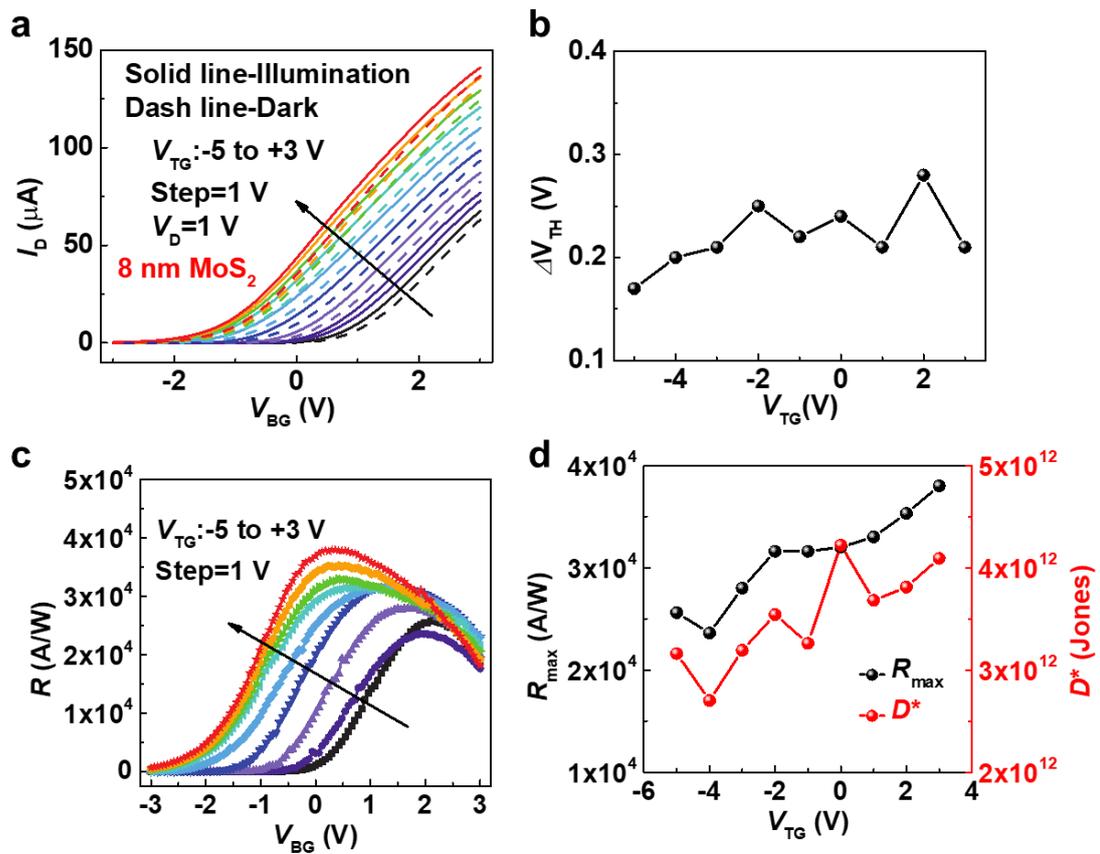

**Figure S7.** Photoresponse characteristic of an 8 nm MoS$_2$ DG phototransistor with different $V_{TG}$. $P_{in}$ is 1.55 mW/cm$^2$ with a wavelength of 550 nm. (a) Transfer characteristics of the devices by sweeping $V_{BG}$ under varied $V_{TG}$ from -5 to +3 V at a step of 1 V under dark (dash line) and illumination (solid line) condition. (b) $V_{TH}$ shift under different $V_{TG}$ due to illumination. (c) $R$ as a function of $V_{BG}$ under different $V_{TG}$. (d) The corresponding $R_{max}$ and $D^*$ as a function of $V_{TG}$.

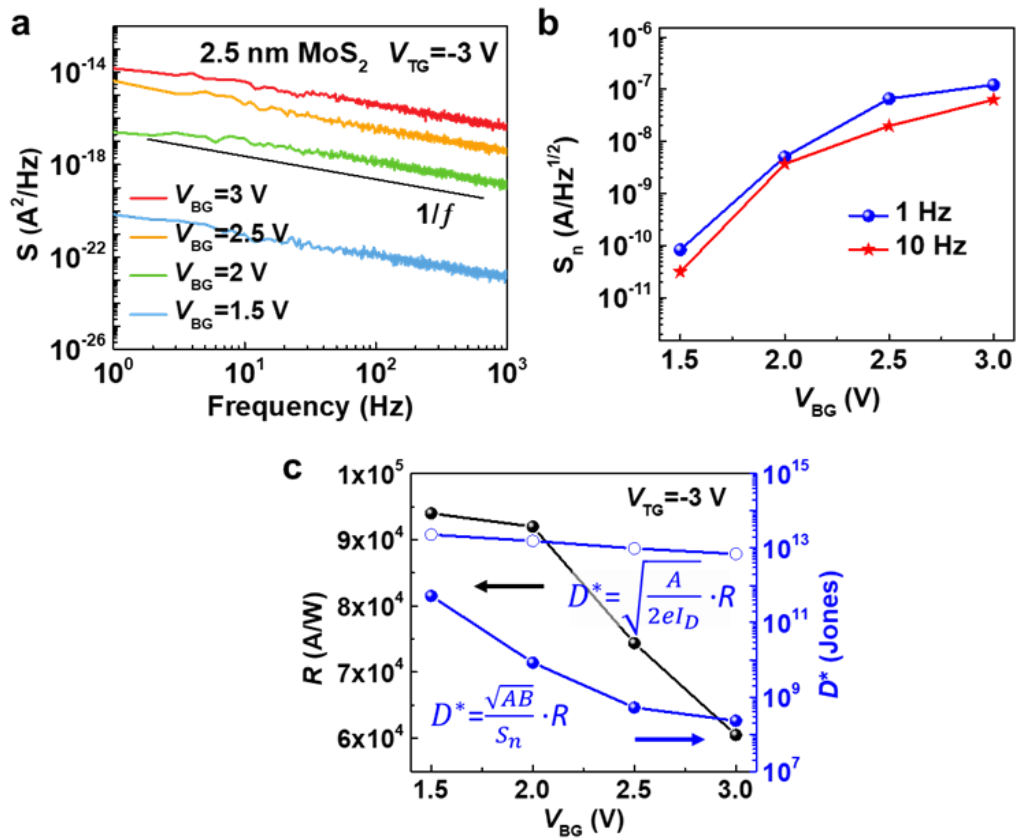

**Figure S8.** Noise analysis of a 2.5 nm MoS$_2$ DG phototransistor at $V_{TG}$=-3 V and $V_D$=1 V. (a) Noise power spectral density (S) for $V_{BG}$=3, 2.5 and 2 V, respectively. (b) Noise current density (S$_n$) extracted from (a) at the noise frequency of 1 and 10 Hz. (c) Responsivity $R$ (black line+black solid sphere) under illumination power density of 1.55 mW/cm$^2$ (550 nm), the measured detectivity $D^*$ by neglecting (blue dash line+blue open circle) and considering 1/$f$-noise (blue line+blue solid sphere).

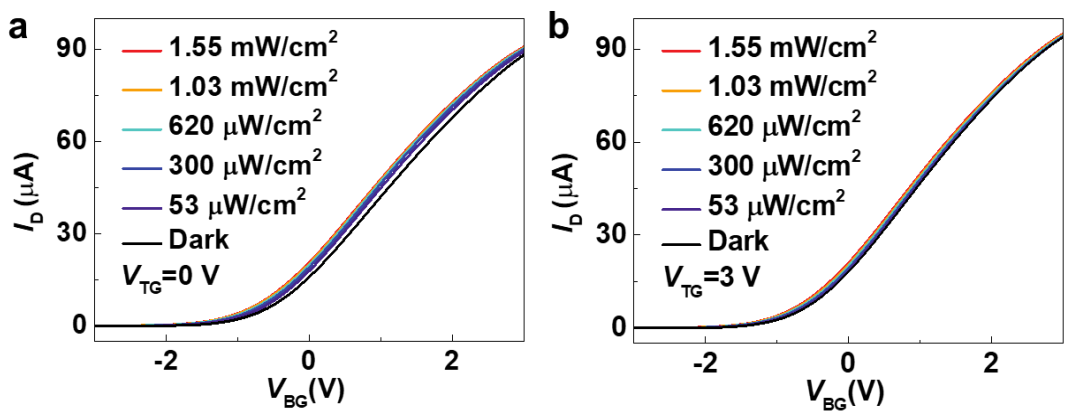

**Figure S9.** Transfer curves under dark and illumination conditions with different $P_{in}$ at $V_{TG}$=0 V (a) and $V_{TG}$=+3 V (b).

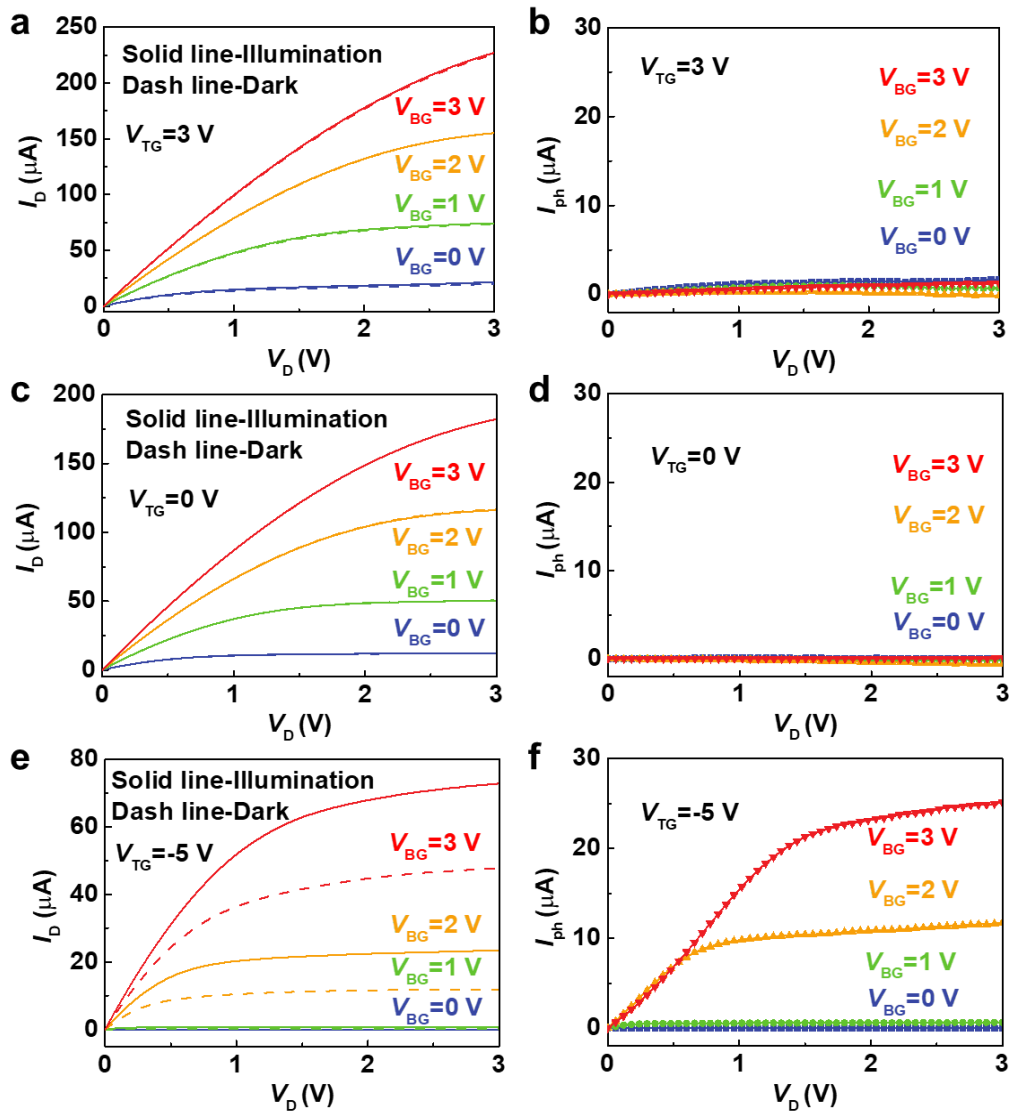

**Figure S10.** Output curves with varying $V_{BG}$ under dark and illumination conditions and its corresponding $I_{ph}$-$V_D$ curves. (a-b) $V_{TG}$=3 V, (c-d) $V_{TG}$=0 V, (e-f) $V_{TG}$=-5 V.

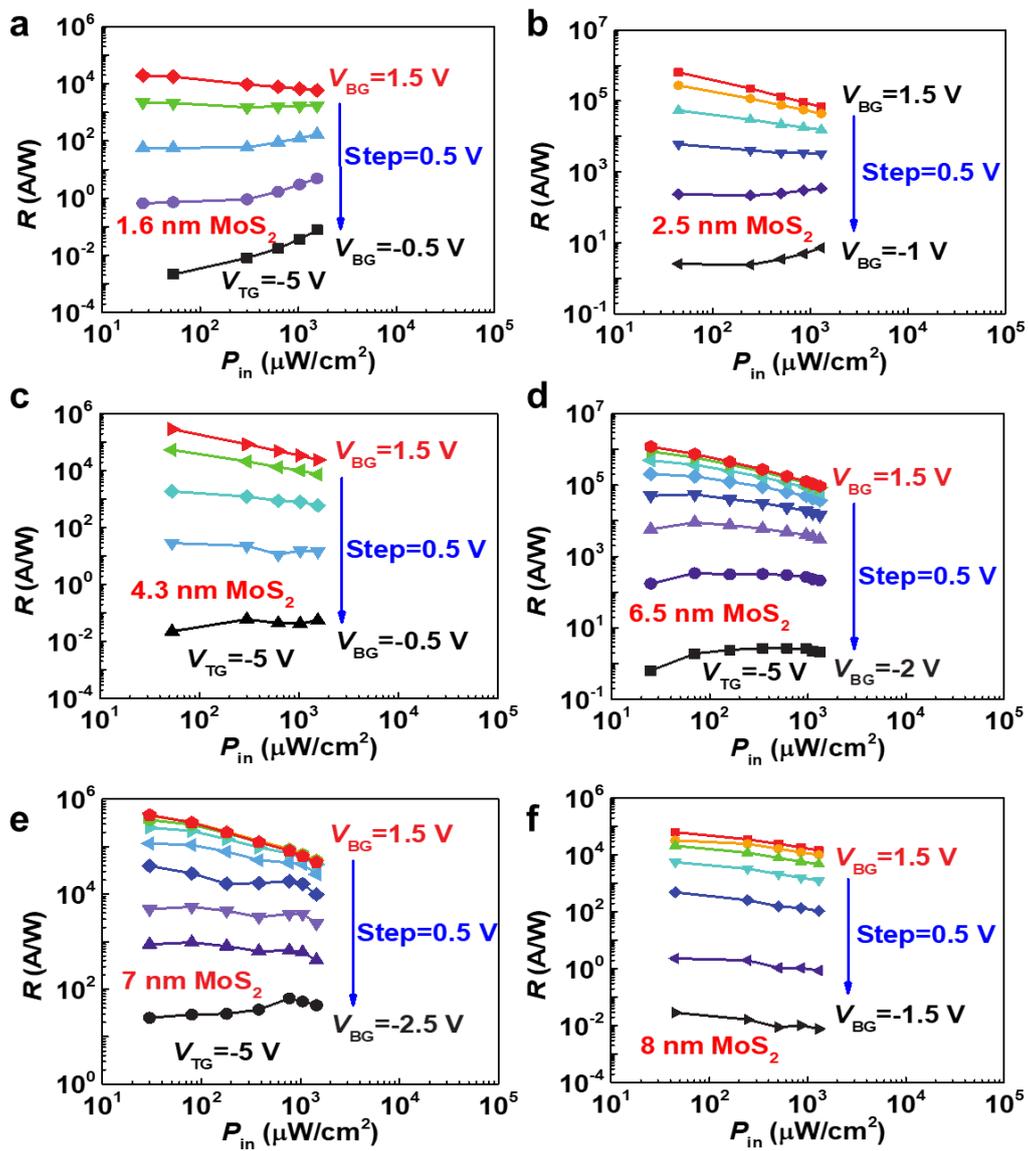

**Figure S11.** The $P_{in}$ dependence of $R$ in different thickness of MoS$_2$ DG phototransistor. (a) 1.6 nm MoS$_2$, (b) 2.5 nm MoS$_2$, (c) 4.3 nm MoS$_2$, (d) 6.5 nm MoS$_2$, (e) 7 nm MoS$_2$ and (f) 8 nm MoS$_2$.

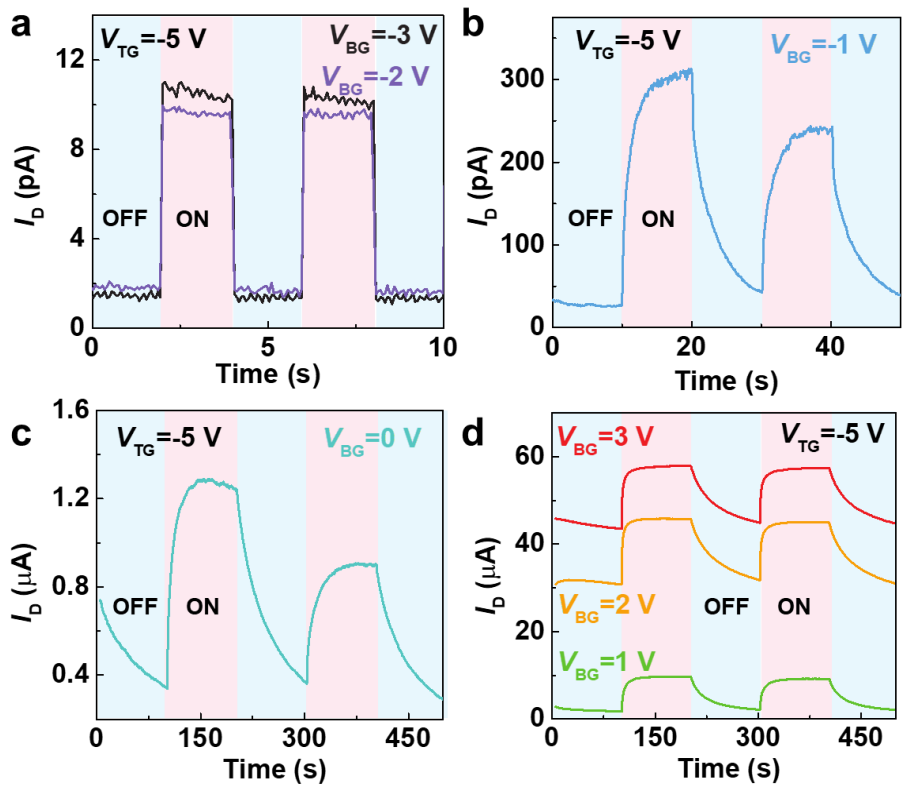

**Figure S12.** Photocurrent pulses of the 5.6 nm MoS$_2$ DG phototransistor measured at 1.55mW/cm$^2$ and $V_{TG}$=-5 V for various $V_{BG}$.

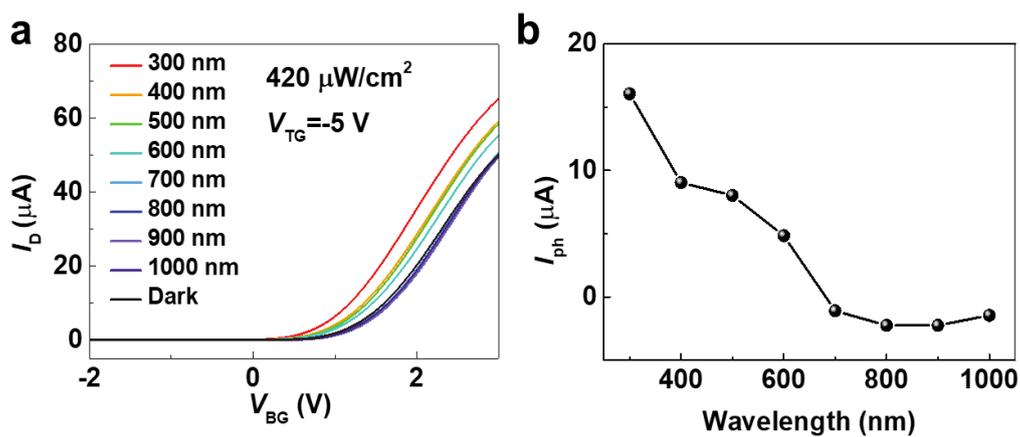

**Figure S13.** The transfer curves of the 5.6 nm MoS$_2$ DG phototransistor with $V_{TG}$=-5 V under dark and illumination of different wavelength (a) and its corresponding $I_{ph}$ as a function of wavelength

**Table S1** Figures-of-merit for $MoS_2$ based photodetectors

| Thickness | Measurement conditions | | | | Responsivity | Detectivity | Response time | Ref. |
|---|---|---|---|---|---|---|---|---|
| L or nm | $V_D$ (V) | $V_G$(V) | $\lambda$ (nm) | $P_{in}$ (mW/cm$^2$) | $R$ (A/W) | $D^*$ (Jones) | $\tau_r/\tau_f$ (ms) | |
| 2.5-6.5 nm | 1 | $V_{TG}$=-5 V, $V_{BG}$=2 V | 550 | $5.3\times10^{-2}$ 1.55 1.55 | $7.7\times10^5$ $7.6\times10^4$ | $1.9\times10^{14}$ $1.9\times10^{13}$ $5\times10^{11}$ | -/- $8.3\times10^3/4.6\times10^4$ | This work |
| 1 L | 1 | 50 | 532 | 80 μW | $7.5\times10^{-3}$ | - | 50/50 | [1] |
| 1 L | 8 | -70 | 561 | $2.4\times10^{-1}$ | $8.8\times10^2$ | - | $4\times10^3/9\times10^3$ | [2] |
| 1 L$^a$ | 1 | 41 | 532 | $1.3\times10^{-1}$ | $2.2\times10^3$ | - | $-/5\times10^5$ | [3] |
| 1 L | 10 | -40 | 532 | $2\times10^1$ | 3.5 | - | -/- | [4] |
| 2 L | 5 | 100 | 532 | $5\times10^1$ | $6.2\times10^3$ | - | $-/2\times10^3$ | [5] |
| 3 L | 10 | 0 | 532 | $2\times10^3$ | $5.7\times10^2$ | $\sim10^{10}$ | $7\times10^2/1\times10^{-1}$ | [6] |
| 5.5 nm | -5 | 0 | 637 | 2.3 | $2.7\times10^4$ | - | $-/0.75\times10^{-1}$ | [7] |
| 3 L | 5 | 0 | 635 | 3.2 | $2.6\times10^3$ | $2.2\times10^{12}$ | 1.8/2 | [8] |
| 1 L | 5 | 0 | 635 | $2.5\times10^{-3}$ | $\sim10^4$ | $7.7\times10^{12}$ | $-/1\times10^4$ | [9] |
| 5 L | 5 | 0 | 850 | - | 1.8 | $5\times10^8$ | 300/360 | [10] |
| 1.3 nm | 10 | 0 | 520 | $5\times10^1$ | $6.3\times10^{-5}$ | $4.2\times10^8$ | -/20 | [11] |
| 2 L | 3 | 0 | 532 | $1\times10^2$ | 0.55 | - | 0.2/1.7 | [12] |
| 80 nm | 1 | 8 | 532 | 2 | 343 | - | -/- | [13] |
| 24 nm | 5 | -35 | 655 | 5.8 | 56.5 | $4.5\times10^9$ | -/- | [14] |
| 16.2 nm | 0.1 | 0 | 454 | $10^{-2}$ | $10^4$-$10^5$ | - | -/- | [15] |
| 56 nm | 1.2 | 0 | 532 | 1.69 | 59 | - | $4.2\times10^{-2}$ | [16] |
| 38 nm | 1 | -30 | 532 | 8.5 | 65.2 | - | $3.5\times10^3/6.7\times10^3$ | [17] |
| 1 L | 10 | 80 | 520 | $7.2\times10^1$ | $10^{-3}$ | $\sim10^7$ | $8\times10^2/1.7\times10^3$ | [18] |
| 1 L$^b$ | 0.1 | 0 | 532 | $1.5\times10^{-1}$ | 99.9 | $9.4\times10^{12}$ | $1.7\times10^4/5.2\times10^3$ | [19] |
| 60 nm$^b$ | 1.5 | 0 | 500 | - | 5.07 | $3\times10^{10}$ | 100-200 | [20] |
| 11 nm$^b$ | 1 | 60 | 655 | $1\times10^2$ | 0.03 | - | -/- | [21] |
| 7.2 nm$^b$ | 10 | 60 | 635 | $7.3\times10^{-8}$ | $7\times10^4$ | $3.5\times10^{14}$ | 20/20 | [22] |
| 3 L$^c$ | 1 | 80 | 405 | 50 nW | $1.6\times10^4$ | - | $70/1.23\times10^3$ | [23] |
| 6 L$^c$ | 1 | -60 | 635 | $3.3\times10^{-3}$ | $6\times10^5$ | $10^{14}$ | $-/\sim350$ | [24] |
| 2 L | 3 | 0 | 532 | 35 pW | $1\times10^5$ | - | $5\times10^5/2\times10^5$ | [25] |

$^a$ Measurement in vacuum. $^b$ Chemical Doping, $^c$ quantum dots, yellow background: $D^*$ extracted by low frequency noise measurements.